\pdfoutput=1
\documentclass[11pt]{article}
\usepackage{jheppub}
\usepackage{epsfig}
\usepackage{amssymb}
\usepackage{amsmath}
\usepackage{amsfonts}
\usepackage{mathrsfs}
\usepackage{hyperref}
\usepackage{multirow}
\usepackage{dsfont}

\def\pb{\bar\partial}

\def\({\left(}
\def\){\right)}
\def\<{\left\langle\,}
\def\>{\, \right\rangle}
\def\[{\left[}
\def\]{\right]}

\def\a{\alpha}
\def\b{\beta}
\def\g{\gamma}
\def\l{\lambda}

\def\t{\theta}

\def\p{\partial}

\def\s{\sigma}

\title{\boldmath $N$-Point Tree-Level Scattering Amplitude\\in the New Berkovits' String}
\author[a,b]{Humberto Gomez}
\author[a,c]{Ellis Ye Yuan}
\affiliation[a]{Perimeter Institute for Theoretical Physics, Waterloo, ON N2L 2Y5,
Canada}
\affiliation[b]{Instituto de F\'isica Te\'orica
UNESP - Universidade Estadual Paulista,\\ Caixa Postal 70532-2
01156-970 S\~ao Paulo, SP, Brazil}
\affiliation[c]{Physics Department, University of Waterloo, Waterloo, ON N2L 2Y5,
Canada}
\emailAdd{humgomzu@ift.unesp.br}
\emailAdd{yyuan@pitp.ca}
\arxivnumber{1312.5485}

\abstract{
We give a proof by direct computation that at tree level, the twistor-like superstring theory in the pure spinor formalism proposed very recently by Berkovits describes ten-dimensional $\mathcal{N}=1$ super Yang-Mills in its heterotic version, and type II supergravity in its type II version. The Yang-Mills case agrees with the result obtained by Mafra, Schlotterer, Stieberger and Tsimpis. When restricting to gluon and graviton scattering, this new theory gives rise to Cachazo-He-Yuan formula.    
}

\begin{document}

    %
    %
    %
    %

\maketitle
\flushbottom

\section{Introduction}

Recently a new formula was proposed by Cachazo, He and Yuan (CHY) to compute the tree-level scattering amplitudes of massless bosons (doubly-colored scalar with cubic self-interaction, pure gluon and pure graviton) in any dimensions~\cite{Cachazo:2013hca,Cachazo:2013iea}, which is constructed upon scattering equations that govern the relation between scattering data and an underlying punctured Riemann sphere in the connected prescription~\cite{Witten:2003nn,Roiban:2004yf,Cachazo:2012da,Cachazo:2012kg,Cachazo:2013zc,Cachazo:2013iaa}. This formula has been proven by Britto-Cachazo-Feng-Witten (BCFW) recursion relations~\cite{Dolan:2013isa}. Given the twistor string origin of such construction, Mason and Skinner found a new ambitwistor string theory whose tree-level scattering produces this formula~\cite{Mason:2013sva}. Moreover, they pointed out that this new version of twistor string can be obtained by taking the chiral infinite tension limit of the ordinary string theory and they gave an explicit example in the bosonic case. This was extended by Berkovits very recently to the superstring in the pure spinor formalism~\cite{nathannewpaper}. By investigating its connection with the RNS formalism in Mason and Skinner's discussion, this twistor-like string theory was also claimed to give rise to the scattering-equation-based formula. A particularly interesting aspect of this extension is that, since the pure spinor formalism naturally encodes space-time supersymmetry, this has the potential of extending the original CHY formula to the supersymmetric case, at least in ten dimensions where this twistor-like string theory sits. So as the first step, it is worth to see how the CHY formula arises from Berkovits' theory in a direct way and how supersymmetry enters into the formula.

In this short paper, we give a direct proof that at tree-level Berkovits' twistor-like theory of heterotic string and type II string is identical to the ten-dimensional $\mathcal{N}=1$ super Yang-Mills (SYM) and type II supergraivty (SUGRA) respectively. The proof parallels the work of Mafra, Schlotterer and Stieberger (and also later on with Broedel) on the disk amplitudes of ordinary superstring in the pure spinor formalism~\cite{npointsmafra,mafratwo,Broedel:2013tta}. This is expected since the constructions of vertex operators are very similar between the two theories. The main difference comes with the moduli, which are now holomorphic coordinates on a Riemann sphere instead of ordered coordinates on the real axis that by conformal symmetry describe points on the boundary of a disk. With this, the scattering equations directly make an appearance in the amplitude~\cite{Mason:2013sva,nathannewpaper}, which is a hint that the formula is of the CHY type. Indeed, the final result obtained here shares a similar Kawai-Lewellen-Tye (KLT) structure with the string amplitude given in \cite{Broedel:2013tta}. And as a result of KLT orthogonality pointed out in \cite{Cachazo:2013iea}, in the heterotic version this just reduces to the SYM amplitude as given in \cite{mafraSYM}, and in the type II version it leads to SUGRA as the KLT of two copies of SYM. Moreover, by applying KLT orthogonality in a different way, this result is exactly equivalent to the original CHY formula when restricting to gluon and graviton scattering.

The paper is organized as follows. We first give a detailed proof for the case of heterotic string in Section \ref{section2}. Since the proof for the type II string shares a lot in common, we only discuss in detail the differences in Section \ref{section3}. A quick review of Berkovits' theory in each case is summarized at the beginning of each section.

\section{Tree-Level SYM Amplitude}\label{section2}

The action in Berkovits' new twistor-like theory for heterotic superstring, which is expected to describe the ${\cal N}=1$ SYM in ten dimensions, is given by~\cite{nathannewpaper}
\begin{equation}\label{action}
S= \int d^2z\, (P_m \pb X^m+ p_\a\pb \t^\a + w_\a\pb \l^\a + b\pb c) + S_c, 
\end{equation}
where $\l^\a$ is a ten dimensional pure spinor (with the constraint $\l\g^m\l=0$) and $S_C$ is the worldsheet action for the current algebra. The BRST operator is defined as
\begin{equation}
Q=\int dz\, (\l^\a d_\a +c(P_m\p X^m p_\a\p \t^\a + w_\a\p \l^\a + T_c)+ bc\p c),
\end{equation}
where $d_\a$ is the Green-Schwarz constraint
\begin{equation}\label{gsc}
d_\a = p_\a - \frac{1}{2}P_m(\g^m\t)_\a.
\end{equation}
The massless vertex operator describing the ${\cal N}=1$ SYM multiplet  are
\begin{equation}\label{vertexym}
\begin{array}{cc}
\begin{split}
V=& c\, \tilde V^I\, J_I,\\U=& \,\tilde U^I \,J_I,
\end{split}&
\begin{split}
&\qquad\text{Unintegrated},\\&\qquad\text{Integrated},
\end{split}
\end{array}
\end{equation}
where
\begin{equation}\label{vertexym2}
\begin{split}
\tilde V^I=& e^{ik\cdot X}\l^\a A^I_\a(\t),\\
\tilde U^I=&e^{ik\cdot X} \bar\delta(K\cdot P)[P^m A^I_m + d_\a W^{\a I} + \frac{1}{2}N_{mn}{\cal F}^{mn I}].
\end{split}
\end{equation}
In the above, $N_{mn}=\frac{1}{2}(\l\g_{mn}w)$, $\{ A_\a,A_m, W^\a,{\cal F}^{mn}\}$ are the ${\cal N}=1$ SYM superfields,  and the current $J_I(\s)$ satisfies
\begin{equation}
J_I(\s_i)J_J(\s_j)\sim \frac{k\delta_{IJ}}{\s^2_{ij}}+ \frac{f_{IJ}^KJ_K(\s_j)}{\s_{ij}}.
\end{equation}

From the action (\ref{action}) it is simple to read the OPE's
\begin{equation}\label{ope}
\begin{split}
&d_\a(\s_i)d_\b(\s_j)\sim -\frac{\g^m_{\a\b}P_m}{\s_{ij}}, \quad d_\a(\s_i)\t^\b(\s_j)\sim \frac{\delta_{\a}^{\b}}{\s_{ij}},\quad
d_\a(\s_i)f(X(\s_j),\t(\s_j))\sim \frac{D_\a f(X(\s_j)}{\s_{ij}},\\
& N^{mn}(\s_i)N_{pq}(\s_j)\sim \frac{4N^{[m}_{\,\,\,\,\,\,[p}\delta^{n]}_{q]}}{\s_{ij}}-\frac{6\delta^{m}_{[p}\delta^{n}_{q]}}{\s^2_{ij}},\quad N^{mn}(\s_i)\l^\a(\s_j)\sim -\frac{1}{2}\frac{(\l\g^{mn})^\a}{\s_{ij}},
\quad P^m(\s_i) P_n(\s_j)\sim 0,\\
&P^m(\s_i) f(X(\s_j),\t(\s_j))\sim -\frac{(k^m)f(X(\s_j),\t(\s_j))}{\s_{ij}},
\end{split}
\end{equation}
where $D_\a:=\p_\a + \frac{1}{2}(\g^m\t)_\a\p_m$ is the covariant derivative. Note that these are the same OPE's as found in the pure spinor superstring formalism \cite{nathanps}, except for the OPE of $P^m(\s_i) P_n(\s_j)$, which in the pure spinor formalism of ordinary string has a double pole\footnote{Note that the OPE between $P^m$ and any superfield has the same convention as in \cite{npointsmafra}. The idea is to apply the results obtained in that paper.}.

The tree-level amplitude prescription is given by the correlation function
\begin{equation}
\mathcal{A}_N= \int \prod_{i=1}^{N-2}d\s_i\, \langle V_1(\sigma_1=0)U_2 \cdots U_{N-2} V_{N-1}(\s_{N-1}=1) V_N(\s_N=\infty)\rangle,
\end{equation}
where the three unintegrated vertex operators $\{V_1(\s_{1}=0),  V_{N-1}(\s_{N-1}=1), V_N(\s_N=\infty)\}$ fix the $SL(2,\mathbb{C})$ gauge symmetry on the sphere. Since there is no correlation between $c$, $J_I$ and the vertices $\{\tilde V^I,\tilde U^I\}$, the above formula can be decomposed as  
\begin{equation}\label{prescription}
\mathcal{A}_N= \int \prod_{i=1}^{N-2}d\s_i
\langle c(\s_1)c(\s_{N-1})c(\s_N) \rangle \,\,\langle \tilde V^{I_1}_1\tilde U^{I_2}_2 ... \tilde U^{I_{N-2}}_{N-2} \tilde V^{I_{N-1}}_{N-1}\tilde V^{I_{N}}_N\rangle\,\,
\langle J_{I_1}J_{I_2} ... J_{I_{N}}\rangle,
\end{equation}
where the $c$-ghost correlator just produces a Vandermonde factor
\begin{equation}
\langle c(\s_1)c(\s_{N-1})c(\s_N) \rangle = \s_{1,N-1}\s_{N-1,N}\s_{N,1}.
\end{equation}

\subsection{$X^m$ and $P^m$ Integration}

We first perform the phase space integration. In the path integral prescription \eqref{prescription} the $X^m$ effective action contribution is given by
\begin{equation}
S[X,P] = -\int d^2\s \left(\frac{1}{2\pi}P_m\pb X^m -i\sum_{i=1}^{N}k_i\cdot X \delta^{(2)}(\s-\s_i)\right).
\end{equation}
Integrating the zero modes of the $X^m$ field leads to the usual momentum conservation $\delta^{(10)}(\sum_i k^m_i)$. The non-zero modes integration implies the constraint 
\begin{equation}
\pb P^m=-2\pi i\sum_{i=1}^{N}k_i^m \delta^{(2)}(\s-\s_i),
\end{equation}
which, on the sphere, has the unique solution
\begin{equation}
P^m(\s)= \sum_{i=1}^N \frac{-(ik_i^m)}{\s-\s_i}.
\end{equation}  
This solution must be replaced in the integrated vertex operators, i.e., at the vertex $U_i$ we have
\begin{equation}\label{Pope}
P^m\,\,\longrightarrow \,\,P^m(\s_i)= \sum_{j\neq i}^N \frac{-(ik_j^m)}{\s_i-\s_j}
\end{equation}
\begin{equation}\label{sceq}
\bar\delta(k_i\cdot P)\,\,\longrightarrow \,\,\bar\delta(k_i\cdot P(\s_i))= \delta\left(i\sum_{j\neq i}^N \frac{(ik_j)\cdot(ik_j)}{\s_i-\s_j}\right)=\delta\left(i\sum_{j\neq i}^N \frac{s_{ij}}{\s_{ij}}\right),
\end{equation}
where\footnote{The definition of $s_{ij}$ matches with one given in \cite{npointsmafra}.} $s_{ij}:=(ik_i)\cdot (ik_j)$. The solution (\ref{Pope}) is equivalent to consider the OPE given in \eqref{ope} and we can write the Dirac delta as 
\begin{equation}\label{scatteringeq}
\bar\delta(k_i\cdot P(\s_i))=\delta\left(\sum_{j\neq i}^N \frac{s_{ij}}{\s_{ij}}\right),
\end{equation}
since the overall $i$ factor does not affect the final answer\footnote{Integrating out the phase space $\{P^m, X^m\}$ implies that it is not necessary to consider the OPE between the Dirac delta $\bar\delta(k\cdot P)$ and the superfields}. Hence, we can conclude that the integration by the $X^m$ and $P^m$ fields imply the  OPE \eqref{ope}, the $N-3$ independent scattering equations \eqref{scatteringeq} and the momentum conservation.

\subsection{Connection to CHY Formula}

In order to compute the pure spinor correlator we must note that every  single pole contraction is the same as those given in \cite{npointsmafra}. This is simple to see, since the only difference between the operators in \eqref{vertexym2} and those used in \cite{npointsmafra} is the missing term 
\begin{equation}\label{missingterm}
\p\t^\a A_\a(X,\t)
\end{equation}   
in the definition of the integrated vertices, whose OPE's involve only double poles (all possible simple poles from this term cancel away in the end). In addition, the operator $\Pi^{m}$ in the ordinary string, which is replaced by $P^m$ in the new vertex operator (\ref{vertexym2}), has the OPE's
\begin{equation}\label{Piope}
\Pi^m(\s_i)\Pi_n(\s_j)\sim -\frac{\delta^m_n}{\s^2_{ij}}, \qquad \Pi^m(\s_i) f(X(\s_j),\t(\s_j))\sim -\frac{k^mf(X(\s_j),\t(\s_j))}{\s_{ij}}
\end{equation}
where $f(X(\s_j),\t(\s_j))$ is any superfield. Note that the only difference  between (\ref{ope}) and (\ref{Piope}) is the double pole. These indicate that all differences enter into the terms with double poles, and so we must have a careful look at these terms before moving on.

In \cite{npointsmafra} it was argued that terms involving double poles always combine to produce a prefactor of the form
\begin{equation}\label{doublepoleprefactor}
\frac{(1+s_{ij})}{\sigma^2_{ij}},
\end{equation} 
whose numerator plays the role of canceling the tachyon pole $1/(1+s_{ij})$ produced by integration of the Koba-Nielsen (KN) factor $\prod_{i<j}|\sigma_i-\sigma_j|^{-s_{ij}}$ (since such a pole is expected to be spurious). At the integrand level this means that the double poles are actually spurious as well, and hence the aim is to dissolve the appearance of the double poles. In the treatment of \cite{npointsmafra} for ordinary string, this is done by integration by parts in the presence of the KN factor, e.g.,
\begin{equation}
\int d\sigma_a\frac{d}{d\sigma_a}\left(\frac{1}{\sigma_{a,b}}\prod_{i<j}|\sigma_i-\sigma_j|^{-s_{ij}}\right)
=-\int d\sigma_a\left(\frac{1+s_{ab}}{\sigma_{ab}^2}+\sum_{i\neq a,b}\frac{s_{ai}}{\sigma_{ab}\sigma_{ai}}\right)\prod_{i<j}|\sigma_i-\sigma_j|^{-s_{ij}}=0,
\end{equation}
and so the effect of this operation is equivalent to the substitution
\begin{equation}\label{substitutionordinarystring}
\frac{(1+s_{ab})}{\sigma^2_{ab}}\longrightarrow-\sum_{i\neq a,b}\frac{s_{ai}}{\sigma_{ab}\sigma_{ai}}.
\end{equation}

It is important to point out that, in the calculation of ordinary string, the presence of the term \eqref{missingterm} in the integrated vertex and the double pole in \eqref{Piope} contribute and only contribute to the term ``$1$'' in the numerator of \eqref{doublepoleprefactor}, and this ``$1$'' term receives no contribution from anything else. Here we just show this explicitly in the simplest example at five points. According to the calculation in \cite{Mafra:2009bz}, when fixing $\{\sigma_1,\sigma_4,\sigma_5\}$, the terms with double poles reads
\begin{equation}\label{doublepoleOPE5pt}
\frac{(1+s_{23})}{\sigma_{23}^2}\,\langle \tilde{V}(\sigma_1)\,[A_\alpha(\sigma_2)W^\alpha(\sigma_3)+A_\alpha(\sigma_3)W^\alpha(\sigma_2)-A_m(\sigma_2)A^m(\sigma_3)]\,\tilde{V}(\sigma_4)\,\tilde{V}(\sigma_5)\rangle,
\end{equation}
where with a slight abuse of notation we denote $\tilde{V}=\lambda_\alpha A^\alpha$. If we study the contribution from the term \eqref{missingterm}, from \eqref{ope} it is easy to see that the only non-trivial OPE's are\footnote{Here we only write out the double-pole terms. As stated before, the simple-pole terms from these additional OPE's eventually cancel each other, and thus are of no interests.}
\begin{equation}\label{vanishing1}
(\partial\theta_\alpha A^\alpha)(\sigma_2)\,(d_\beta W^\beta)(\sigma_3)\sim\frac{A_{\alpha}(\sigma_2)W^\alpha(\sigma_3)}{\sigma_{23}^2},\quad
(d_\beta W^\beta)(\sigma_2)\,(\partial\theta_\alpha A^\alpha)(\sigma_3)\sim\frac{A_{\alpha}(\sigma_3)W^\alpha(\sigma_2)}{\sigma_{23}^2}.
\end{equation}
Moreover, the non-trivial OPE among $\Pi_m$'s given in \eqref{Piope} produces an additional term
\begin{equation}\label{vanishing2}
(\Pi^mA_m)(\sigma_2)\,(\Pi^nA_n)(\sigma_3)\sim-\frac{A_m(\sigma_2)A^m(\sigma_3)}{\sigma_{23}^2}.
\end{equation}
When we switch from ordinary string to the twistor string constructed by Berkovits, one can check that \eqref{vanishing1} and \eqref{vanishing2} are the only OPE's that cease to contribute to the vertices correlator, and so the change to the result \eqref{doublepoleOPE5pt} is only to delete the ``$1$'' from the prefactor $(1+s_{23})$. In general, in the computation of Berkovits' twistor string, we just need to switch the prefactors $(1+s_{ij})$ to the corresponding $s_{ij}$~\footnote{The authors are grateful to Carlos Mafra for discussions over this issue.}.

Now, in the context of twistor string, there is no longer any KN factor. Instead, since the $\sigma$ variables are evaluated under the delta constraints \eqref{scatteringeq}, the way to dissolve the presence of double poles is to apply substitution on the support of the corresponding scattering equations, e.g.,
\begin{equation}\label{substitutiontwistorstring}
\frac{s_{ab}}{\sigma^2_{ab}}\longrightarrow-\sum_{i\neq a,b}\frac{s_{ai}}{\sigma_{ab}\sigma_{ai}}.
\end{equation}
From \eqref{substitutionordinarystring} and \eqref{substitutiontwistorstring}, we see that although the differences in OPE's between ordinary string and Berkovits' twistor string lead to different appearances of double-pole terms, after canceling these spurious poles they actually give the same result for the vertices correlator. 

Due to this fact, we are justified to directly apply the results obtained in \cite{npointsmafra}
\begin{equation}\label{correlatorpsUV}
\begin{split}
\langle \tilde V^{I_1}_1\tilde U^{I_2}_2 ... \tilde U^{I_{N-2}}_{N-2} \tilde V^{I_{N-1}}_{N-1}\tilde V^{I_{N}}_N\rangle
=&\delta^{(10)}(\sum_i k^m_i)\,\prod_{i=2}^{N-2}\delta\left(\sum_{j\neq i}^N \frac{s_{ij}}{\s_{ij}}\right)\cdot\\
&\cdot\sum_{\b\in S_{N-3}}A_{YM}(1,\b,N-1,N)
\,\,\prod_{k=2}^{N-2}\sum_{m=1}^{k-1}\frac{s_{\b(m)\b(k)}}{\sigma_{\b(m)\b(k)}},
\end{split}
\end{equation}
where $A_{YM}(1,\b,N-1,N)=A_{YM}(1,\b(2),...,\b(N-3),N-1,N)$ is the SYM scattering amplitude  which is given in terms of the BRST building blocks \cite{mafraSYM}.
Furthermore, from \cite{Broedel:2013tta} we know that by manipulations with partial fraction relations the last factor above can be rewritten as\footnote{Rigorously speaking, this identity holds only when $\sigma_N$ is gauge-fixed at infinity. However, for the general gauge where $\sigma_N$ is finite, the requirement of $SL(2,\mathbb{C})$ invariance of $\mathcal{A}_N$ guarantees that the r.h.s.~below is the correct answer. This will become obvious later in \eqref{measure}.}
\begin{equation}\label{resumKLT}
\prod_{k=2}^{N-2}\sum_{m=1}^{k-1}\frac{s_{\b(m)\b(k)}}{\sigma_{\b(m)\b(k)}}
=\sigma_{1,N}\sigma_{N,N-1}\sigma_{N-1,1}\sum_{\gamma\in S_{N-3}}\mathcal{S}[\beta|\gamma]\frac{1}{(1,\gamma,N,N-1)},
\end{equation}
where
$$ 
(1,\gamma,N,N-1):= \s_{1\gamma(2)}\s_{\gamma(2)\gamma(3)}...\s_{\gamma(N-2)N}\s_{N,N-1}\s_{N-1,1}
$$
denotes the Parker-Taylor factor, and
$$
\mathcal{S}[\beta|\gamma]:=\prod_{a=1}^{n-2}\left(s_{1,\beta(a)}+\sum_{b=2}^{a-1}\theta(\beta(b),\beta(a))_{\gamma}\, s_{\beta(b),\beta(a)}\right)
$$
is the $(n-3)!\times(n-3)!$ based Kawai-Lewellen-Tye (KLT) kernel, with $\theta(a,b)_{\beta}=1$ if the ordering of the labels $a,b$ is the same in both orderings $\beta$ and $\gamma$, and zero otherwise~\cite{BjerrumBohr:2010ta}.

On the other hand, the current algebra correlator gives\footnote{In addition to the single-trace terms, the current algebra correlator also produces multi-trace terms \cite{Berkovits:2004hg,Mason:2013sva}. As stated in \cite{Mason:2013sva}, the multi-trace terms are associated to coupling Yang-Mills to gravity. Here we care about pure Yang-Mills and so we only focus on the single-trace terms.}
\begin{equation}\label{correlatorpsJ}
\langle J_{I_1}J_{I_2}\cdots J_{I_{N}}\rangle=\sum_{\Pi\in S_{N-1}}\frac{\text{Tr}(T^{I_1} T^{\Pi(I_2)}\cdots T^{\Pi(I_N)})}{(1,\Pi(2),\ldots,\Pi(N))}.
\end{equation}
Due to the delta constraints in \eqref{correlatorpsUV}, the formula actually reduces to a rational function with the $\{\sigma\}$ variables evaluated on the solutions to the scattering equations. On the support of these equations, it is known from \cite{Cachazo:2012uq} that the Parke-Taylor factors in \eqref{correlatorpsJ} can be linearly decomposed onto a $(n-3)!$ basis due to the validity of Bern-Carrasco-Johansson relations~\cite{Bern:2008qj}
\begin{equation}
\frac{1}{(1,\Pi(2),\ldots,\Pi(N))}=\sum_{\alpha\in S_{N-3}}\mathcal{K}[\Pi,\alpha]\frac{1}{(1,\alpha,N-1,N)}
\end{equation}
in the same way as
\begin{equation}\label{YMdecomposition}
A_{YM}(1,\Pi(2),\ldots,\Pi(N))=\sum_{\alpha\in S_{N-3}}\mathcal{K}[\Pi,\alpha]A_{YM}(1,\alpha,N-1,N),
\end{equation}
with $\mathcal{K}[\Pi,\alpha]$ some function only depending on the kinematic invariants $\{s_{ij}\}$ and the two orderings $\Pi,\alpha$ (which is not relavent to our discussion).

To this end, we see that the two copies of Vandermonde factor $(\s_{1,N-1}s_{N-1,N}s_{N,1})$ from the $c$-ghost correlation and \eqref{resumKLT} combine with the measure and the delta constraints in \eqref{correlatorpsUV} to form fully permutation invariant and $SL(2,\mathbb{C})$ covariant objects
\begin{equation}\label{measure}
\begin{split}
\int\frac{d^n\s}{\text{vol }SL(2,\mathbb{C})}&:=\s_{1,N-1}\s_{N-1,N}\s_{N,1}\int\sum_{i-2}^{N-2}d\s_i,\\
{\prod}'\left(\sum\frac{s_{ij}}{\s_{ij}}\right)&:=\s_{1,N}\s_{N,N-1}\s_{N-1,1}\prod_{i=2}^{N-2}\delta\left(\sum_{j\neq i}^N \frac{s_{ij}}{\s_{ij}}\right).
\end{split}
\end{equation}
Hence by assembling different pieces in \eqref{prescription}, the whole amplitude can be expressed as
\begin{equation}\label{fullYMamplitudeintermediate}
\begin{split}
\mathcal{A}_N
=&
\sum_{\Pi\in S_{N-1}}\sum_{\alpha\in S_{N-3}}\text{Tr}(T^{I_1} T^{\Pi(I_2)}\cdots T^{\Pi(I_N)})\mathcal{K}[\Pi,\alpha]\int\frac{d^n\s}{\text{vol }SL(2,\mathbb{C})}{\prod}'\left(\sum\frac{s_{ij}}{\s_{ij}}\right)\frac{1}{(1,\alpha,N-1,N)}\\
&\sum_{\b\in S_{N-3}}A_{YM}(1,\b,N-1,N)\sum_{\gamma\in S_{N-3}}\mathcal{S}[\beta|\gamma]\frac{1}{(1,\gamma,N,N-1)}\,\delta^{(10)}(\sum_i k^m_i).
\end{split}
\end{equation}
It is easy to see that the part
\begin{equation}\label{mdef}
m[\gamma|\alpha]:=\int\frac{d^n\s}{\text{vol }SL(2,\mathbb{C})}{\prod}'\left(\sum\frac{s_{ij}}{\s_{ij}}\right)\frac{1}{(1,\gamma,N,N-1)}\frac{1}{(1,\alpha,N-1,N)}
\end{equation}
is exactly the double partial amplitude in the doubly colored $\phi^3$ theory computed by CHY formula in \cite{Cachazo:2013iea}. Since from there we know that as the result of KLT orthogonality
\begin{equation}\label{mrelations}
m[\gamma|\alpha]=(\mathcal{S}[\gamma|\alpha])^{-1},
\end{equation}
\eqref{fullYMamplitudeintermediate} reduces to
\begin{equation}
\mathcal{A}_N
=\delta^{(10)}(\sum_i k^m_i)\,\sum_{\Pi\in S_{N-1}}\sum_{\alpha\in S_{N-3}}\text{Tr}(T^{I_1} T^{\Pi(I_2)}\cdots T^{\Pi(I_N)})\mathcal{K}[\Pi,\alpha]A_{YM}(1,\alpha,N-1,N),
\end{equation}
which by \eqref{YMdecomposition} is indeed the full tree-level amplitude of ten-dimensional $\mathcal{N}=1$ SYM as originally computed in \cite{mafraSYM}. Supersymmetries are solely encoded into the $(n-3)!$ basis $A_{YM}(1,\alpha,N-1,N)$.

On the other hand, for the component amplitude involving gluons only, if we substitute the $A_{YM}(1,\b,N-1,N)$ in \eqref{fullYMamplitudeintermediate} by the CHY formula for gluons, when we apply the original KLT orthogonality as stated in~\cite{Cachazo:2013gna}, the factor
\begin{equation}\label{pfaffianpart}
\sum_{\b\in S_{N-3}}A_{YM}(1,\b,N-1,N)\sum_{\gamma\in S_{N-3}}\mathcal{S}[\beta|\gamma]\frac{1}{(1,\gamma,N,N-1)}
\end{equation}
becomes a Pfaffian and the entire \eqref{fullYMamplitudeintermediate} again reduces to the original CHY formula (apart from the momentum conservation).

\section{Tree-Level SUGRA Amplitude}\label{section3}

In the version of Berkovits' theory for type II superstring, which is expected to describe type II supergravity in ten dimensions, the action reads \cite{nathannewpaper} 
\begin{equation}\label{actiontypetwo}
S= \int d^2z (P_m \pb X^m+ p_\a\pb \t^\a + w_\a\pb \l^\a + \hat p_{\hat \a}\pb \hat \t^{\hat \a} + \hat w_{\hat \a}\pb \hat \l^{\hat \a}), 
\end{equation}
where $\l^\a$ and $\hat \l^{\hat \a}$ are pure spinors. The BRST charge is defined as
\begin{equation}
Q=\int dz (\l^\a d_\a +\hat \l^{\hat \a}\hat  d_{\hat \a})
\end{equation}
where $d_\a$($\hat d_{\hat \a}$) is the Green-Schwarz constraint given in (\ref{gsc}). 

The massless vertex operators are the double copy of the vertices defined previously in (\ref{vertexym}), but now without $c$-ghost and $J^I$ current. With a little change of notation for later convenience, these are given by
\begin{equation}\label{vertexgrv}
\begin{array}{cc}
\begin{split}
V=& \,e^{ik\cdot X} \tilde V\,\, \tilde{\hat{V}},\\U=& \,e^{ik\cdot X} \bar\delta(K\cdot P)\,\tilde U\,\,\tilde{\hat{U}},
\end{split}&
\begin{split}
&\qquad\text{Unintegrated},\\&\qquad\text{Integrated},
\end{split}
\end{array}
\end{equation}
where
\begin{equation}
\begin{split}
\tilde V=& \l^\a A_\a(\t),\\
\tilde U=&P^m A_m + d_\a W^{\a} + \frac{1}{2}N_{mn}{\cal F}^{mn},
\end{split}
\end{equation}
and the $\{\tilde{\hat{V}},\,\tilde{\hat{U}}\}$ are defined in a similar way (with the hatted version of the fields).

\subsection{Connection to CHY Formula}

The computation in this case greatly resembles that for the heterotic string, and so here we only summarize the differences. Since there is no correlation between the hatted and non-hatted fields, the tree-level amplitude prescription reads
\begin{align}\label{scatteringgrv}
\mathcal{M}_N &= \int \prod_{i=1}^{N-2}d\s_i\, \langle V_1(\sigma_1=0)U_2 \cdots U_{N-2} V_{N-1}(\s_{N-1}=1) V_N(\s_N=\infty)\rangle\nonumber\\
&=\delta^{(10)}(\sum_i k^m_i)\,
\int \prod_{i=1}^{N-2}d\s_i\,\,\delta\left(\sum_{j\neq i}^{N}\frac{s_{ij}}{\s_{ij}}\right)
\,\,\langle \tilde V_1 \tilde U_2 ... \tilde U_{N-2} \tilde V_{N-1}\tilde V_N \rangle\,\,\langle \tilde{\hat{V}}_1\tilde{\hat{U}}_2 ... \tilde{\hat{U}}_{N-2} \tilde{\hat{V}}_{N-1}\tilde{\hat{V}}_N\rangle
\end{align}
where we have already performed the phase space integration, which is the same as that discussed in the SYM case. Each of the remaining correlators above is computed in the same way as that in \eqref{correlatorpsUV} and \eqref{resumKLT}. Hence one can check that ${\cal M}_N$ can be expressed as
\begin{equation}\label{sugraformula}
{\cal M}_N=\delta^{(10)}(\sum_i k^m_i)\,
\sum_{\b\in S_{N-3}}\sum_{\hat \b\in S_{N-3}} A_{YM}(1,\b,N-1,N)\,\, H[\b|\hat\b]\,\,\hat A_{YM}(1,\hat \b,N,N-1),
\end{equation}
where
\begin{equation}
\begin{split}
H[\b|\hat\b] =&\int\frac{d^n\s}{\text{vol }SL(2,\mathbb{C})}\,{\prod}'\left(\sum\frac{s_{ij}}{\s_{ij}}\right)
\sum_{\gamma\in S_{N-3}}\mathcal{S}[\beta|\gamma]\frac{1}{(1,\gamma,N,N-1)}
\sum_{\hat{\gamma}\in S_{N-3}}\mathcal{S}[\hat{\gamma}|\hat{\beta}]\frac{1}{(1,\hat{\gamma},N-1,N)}\\
=&\sum_{\gamma,\hat{\gamma}\in S_{N-3}}\mathcal{S}[\beta|\gamma]m[\gamma|\hat{\gamma}]\mathcal{S}[\hat{\gamma}|\hat{\beta}].
\end{split}
\end{equation}
Then by the relation \eqref{mrelations} it is clear that
\begin{equation}
{\cal M}_N=\delta^{(10)}(\sum_i k^m_i)\,
\sum_{\b\in S_{N-3}}\sum_{\hat \b\in S_{N-3}} A_{YM}(1,\b,N-1,N)\,\, \mathcal{S}[\b|\hat\b]\,\,\hat A_{YM}(1,\hat \b,N,N-1),
\end{equation}
which is just the KLT relation in constructing SUGRA amplitude from the corresponding SYM amplitude. So we have also confirmed that this theory indeed describes the type II SUGRA at tree level. In the same manner, when restricting to graviton scattering, if in \eqref{sugraformula} we substitute each of the two SYM amplitudes by their CHY formula, a direct simplification recovers the original CHY formula for pure gravitons.

\section{Discussion}

In this paper we have given a proof by direct computation that at tree level Berkovits' new twistor-like theory for the heterotic string describes the ten-dimensional $N=1$ SYM and that for the type II string describes the corresponding SUGRA. The computation straightforwardly leads to the scattering-equation-based structure in CHY formula, in similar way as discussed by Mason and Skinner in the RNS formalism. Although the result is manifestly supersymmetric, due to the similar structure with the superstring amplitude given in \cite{Broedel:2013tta}, the supersymmetric structure is completely absorbed into the $(n-3)!$ based $A_{YM}$ factors therein. For this reason, it is not at all obvious whether in SYM there exists any compact integrand which is the supersymmetric analog of the Pfaffian factor as appearing in the CHY formula for pure gluon scattering (at least in ten dimensions). However, by comparing \eqref{fullYMamplitudeintermediate} with the CHY formula, we know that if there is such a structure, it has to be equivalent to \eqref{pfaffianpart}. So it would be very nice if one can find a non-trivial way to manipulate \eqref{pfaffianpart} to make manifest such structure embedded therein.

At the time when this paper was being prepared, Adamo, Casali and Skinner published a new work studying Mason and Skinner's ambitwistor string at one loop~\cite{Adamo:2013tsa}. In particular, the extention of scattering equations to loop levels was proposed, and one-loop amplitudes for NS-NS external states in the type II ambitwistor string were calculated. It is interesting to see how Berkovits' theory works at loop levels.

Moreover, here we have started from the action of a twistor-like string theory, which as claimed in \cite{nathannewpaper} is the chiral infinite tension limit of the action for the original string theory. It is still interesting to see how this chiral limit can be explicitly performed at the amplitude level. The usual way of taking the infinite tension limit of string amplitude cannot be done smoothly, i.e., a direct $\alpha'$-expansion of the integrand (mainly the Koba-Nielsen factor) is not allowed, since it is the singularities from the integration boundaries that are responsible for the field theory counterpart. However, given the analogous structure between CHY formula and that of string amplitude obtained in \cite{npointsmafra,mafratwo,Broedel:2013tta}, it is speculated that there might exit a way to smoothly connect the two sides via a novel way of taking the infinite tension limit. If this is true, this chiral limit might have the potential to play the role.

\acknowledgments
The authors would like to thank Nathan Berkovits, Freddy Cachazo and Carlos Mafra for useful discussions, especially Carlos Mafra for pointing out the importance of the double poles from OPE in the calculation. H.G is grateful to the Perimeter Institute for Theoretical Physics for warm hospitality during stages of this work. The work of H.G is supported by FAPESP grant 07/54623-8, 13/11409-7.
This work is supported by Perimeter Institute for Theoretical Physics. Research at Perimeter Institute is supported by the Government of Canada through Industry Canada and by the Province of Ontario through the Ministry of Research \& Innovation.

\bibliographystyle{JHEP}
\bibliography{BerkovitsString}

\end{document}